\documentclass{aastex631}

\usepackage{graphicx}
\usepackage{bm}
\usepackage{xcolor}
\usepackage{amsmath}

\usepackage{hyperref}
\hypersetup{
    colorlinks=true,
    linkcolor=blue,
    filecolor=blue,      
    urlcolor=blue,
    pdftitle={main},
    pdfpagemode=FullScreen,
    }

\begin{document}


\title{Limiting Rotation Rate of Neutron Stars from Crust Breaking and Gravitational Waves}

\author{J.A. Morales}
\email{jormoral@iu.edu}
\affiliation{Center for the Exploration of Energy and Matter and Department of Physics, Indiana University, Bloomington, IN 47405, USA }

\author{C.J. Horowitz}
 \email{horowit@iu.edu}
\affiliation{Center for the Exploration of Energy and Matter and Department of Physics, Indiana University, Bloomington, IN 47405, USA }

\date{\today}

\begin{abstract}
Neutron stars are not observed to spin faster than about half their breakup rate.  This limiting rotational frequency may be related to the strength of their crusts.  As a star spins up from accretion, centrifugal forces stress the crust.  We perform finite-element simulations of rotating neutron stars and find that the crust fails at rotation rates about half the breakup rate.  Given uncertainties in microphysics, we have not determined the crust configuration after this failure.  Instead, we argue that the crust may fail in an asymmetric way and could produce a configuration with a significant ellipticity (fractional difference in moments of inertia).  If the ellipticity is large, a rotating star will radiate gravitational waves that may limit further spin up.  These stars may be promising sources for LIGO / VIRGO and next generation gravitational wave detectors.
\end{abstract}



\section{Introduction}

Neutron stars can gain angular momentum from accretion and spin rapidly \citep{1982CSci...51.1096R}.  Indeed, there are many observed millisecond pulsars \citep{2005AJ....129.1993M}.  However, no neutron star (NS) is observed to spin faster than about half the Keplerian breakup rate \citep{2006Sci...311.1901H}.   Torques from gravitational wave (GW) radiation could limit the rotation rate \citep{Bildsten1998}.  It is unclear what makes a NS asymmetric so that it radiates GW. Some possibilities include asymmetric electron capture reactions \citep{Bildsten1998,10.1046/j.1365-8711.2000.03938.x} or thermal expansion \citep{jones2024neutronstarmountainssupported}.  However, the produced asymmetry may not be large enough to explain the limiting spin \citep{jones2024neutronstarmountainssupported}.  In addition to GW radiation, NS spin up may be limited by properties of the accretion disk-NS magnetosphere system \citep{10.1046/j.1365-8711.2000.03938.x,10.1111/j.1365-2966.2005.09167.x,Haskell_2011} or electromagnetic winds \citep{Parfrey_2016}. These effects may be too small to explain the limiting spin without GW radiation \citep{10.1093/mnras/stad2036}.    

There are many ongoing and near future searches for GW from rotating NS, see for example \citep{Riles2023,Pagliaro2023}.  A number of these searches focused on the X-ray bright accreting system Scorpius X-1 \citep{PhysRevD.95.122003,Abbott2017_4,Abbott2019_3,Abbott2022_3}.  Many searches did not reach the sensitivity to detect a signal if one assumes torques from GW radiation and accretion balance each other.  Recent searchs did reach a torque balance limit but only for some values of unknown parameters such as the spin frequency \citep{Zhang2021,Abbott_2022new}.  No continuous GW signals have yet been detected. Future searches, with both existing and next generation detectors such as Cosmic Explorer \citep{reitze2019cosmicexploreruscontribution}, could be very promising.

In this letter we postulate that the maximum rotation rate of a NS is simply related to the breaking of its crust.   Axisymmetric stars will not radiate GW and can spin up.  However, at some point centrifugal forces will break the crust.   We assume that when the crust breaks it does so asymmetrically.  The released elastic stresses will produce an asymmetric NS that radiates GW.  {\it NS may spin up until their crust breaks.} After that, GW radiation, from the deformed broken crust, may keep the star from spinning faster.

The rotation rate, when the crust breaks, depends on the strength of the crust.  Many early works assumed the crust was weak and could break often.  These breaks could produce star quakes and possibly explain some glitches in the rotation frequency of pulsars \citep{BAYM1971816,1976ApJ...203..213R,1991ApJ...382..587R}.   Molecular dynamics simulations find that the crust is very strong with a breaking strain (fractional deformation) of order 0.1 \citep{PhysRevLett.102.191102}.  This strong crust may significantly change the picture.  Crust breaking may now be rare.  In an unpublished preprint, we estimated, using a simple constant density model, that the crust does not break until the rotation rate is high \citep{Fattoyev2018}. In this letter we perform finite-element simulations, including a large range of crust densities, to more accurately determine when centrifugal forces break the crust.

We first describe our finite-element formalism to calculate rotational perturbations of a Newtonian star with a polytrope equation of state.  We then present results for the displacement and strain of the crust and determine the rotational speed when the maximum strain exceeds the breaking strain.  Finally, we discuss what might happen when the crust breaks and possible implications for gravitational wave observations.  

\section{Formalism}

We begin with a brief summary of our finite-element formalism  \citep{Morales:2024zka} to obtain the structure of the crust of a Newtonian star with a polytropic equation of state.\footnote{Although we use the same formalism,  \cite{Morales:2024zka} was focused on a different problem, how a small intrinsic asymmetry in the elastic properties of the crust produces an asymmetry in the shape of a NS.}  The first step consists of obtaining the zeroth-order stellar structure of a non-rotating, cold, barotropic, and Newtonian NS at mechanical equilibrium:
\begin{equation}
m^{\prime} = 4 \pi r^2 \rho \label{eqn:SEa}
\end{equation}
\begin{equation}
p^{\prime} = -\rho \Phi^{\prime} \label{eqn:SEb}
\end{equation}
\begin{equation}
\Phi^{\prime} = \frac{Gm}{r^2} \label{eqn:SEc} \ .
\end{equation}
Here $r$ is the radial coordinate, $\rho$ is the mass density, $p = K \rho^2$ is the pressure, $K$ is the polytropic proportionality constant, $\Phi$ is the gravitational potential, and $m$ is the enclosed mass. The prime ($^{\prime}$) represents radial derivatives. When the star cools it crystallizes to form the solid crust with an angled-average shear modulus given by $\mu(\rho) \approx \kappa \rho$, where  $\kappa = 10^{16}$ cm$^2$ s$^{-2}$ \citep{Haskell2006}. For a fixed composition, the shear modulus of a Coulomb solid is $\propto \rho^{4/3}$.  However given changes in composition with density, $\kappa\rho$ provides a reasonable fit.  We do not expect much sensitivity to the exact form of $\mu(\rho)$.  We assume that the bottom of the crust is located where the mass density is $\rho_{bottom} = 2 \times 10^{14}$ g/cm$^3$ while the top of the crust is located where the mass density $\rho_{top} = 10^{12}$ g/cm$^3$. As we explained in our previous paper \citep{Morales:2024zka}, placing the top of the crust at this high density does not affect the main results of the simulations and helps to avoid numerical difficulties.

When the NS rotates, the centrifugal acceleration can be treated as a first-order perturbation against the background structure described  by equations (\ref{eqn:SEa}-\ref{eqn:SEc}), as long as the rotational speed of the NS is small. The structural changes that stem from spinning up or down the NS can be described by Eulerian  perturbations $\delta G_{tot}(\vec{r}) = \sum_{l} \delta G_{l}(r) P_l(\text{cos} \ \theta)$, where $G_{tot}$ is a background (zeroth-order) quantity, $\vec{r}$ is the position of a matter element with respect to the star's center, $P_l(\text{cos} \ \theta)$ is the $l$-th Legendre polynomial, and $\theta$ is the polar angle with respect to the rotation axis.

The small centrifugal acceleration leads to the perturbed Euler and Laplace equations
\begin{equation}
    \nabla_i \delta p_{tot} + \delta \rho_{tot} \nabla_i \Phi + \rho \nabla_i \delta \Phi_{tot} - f_i = 0
\label{eqn:pert_euler}
\end{equation}
and
\begin{equation}
    \nabla^2 \delta \Phi_{tot} = 4 \pi G \delta \rho_{tot},
    \label{eqn:pert_poisson}
\end{equation}
where $f_i = \rho \omega^2_{\text{diff}} s \hat{s}$ is the centrifugal force, $s$ is the distance from the axis of rotation, $\hat{s}$ is the corresponding unit vector, $\omega^2_{\text{diff}} \equiv \omega^2 - \omega_0^2$, and $\omega$ and $\omega_0$ are the final and initial angular frequencies, respectively. The small change in pressure is $\delta p_{tot} = c_s^2 \delta \rho_{tot}$, where $c_s$ is the speed of sound. It is convenient to re-write the centrifugal force as $f_i = - \rho \nabla_i \delta \chi_{tot}$. The centrifugal potential $\delta \chi_{tot} \equiv - \tfrac{1}{2} \omega^2_{\text{diff}} s^2$. This definition allows us to encapsulate the perturbed gravitational potential and the centrifugal potential into the variable $\delta U_{tot} \equiv \delta \Phi_{tot} + \delta \chi_{tot}$. Decomposition into Legendre polynomials and separation of variables applied to Eqs. (\ref{eqn:pert_euler}) and (\ref{eqn:pert_poisson}) yield the radial equation
\begin{equation}
    \delta U_l^{\prime \prime} + \frac{2}{r} \delta U_l^{\prime} - \frac{\beta^2}{r^2} \delta U_l = 4 \pi G \delta \rho_l - 2 \omega^2_{\text{diff}} \delta_{l0} \ ,
    \label{eqn:pert_poisson_l}
\end{equation}
where $l$ is the Legendre-polynomial mode, $\beta^2 \equiv l(l+1)$ and $\delta_{l0}$ is 1 if $l=0$ and 0 otherwise. The angular part of each $l$-th Legendre mode of the perturbed Euler equation gives
\begin{equation}
    \delta \rho_l = - \frac{\rho}{c_s^2} \delta U_l \ .
    \label{eqn::pert_density_l}
\end{equation}
For each Legendre mode, the two boundary conditions that are needed to obtain $\delta U_l$ are the regularity of the perturbed gravitational potential at the origin and the interior-exterior matching of the perturbed gravitational potential. Since the latter boundary condition is implicit, the shooting method is used to solve for $\delta U_l$ in each l-th Legendre mode. With this scheme, the perturbed gravitational potential $\delta \Phi_{tot}$ is obtained. Then, we use that perturbed gravitational potential within the equations of hydro-elastic equilibrium to obtain the structure of the deformed crust using the finite-element method. Since the crustal mass is roughly 1 \% of the star, we do not expect the crust to influence the perturbed gravitational potential significantly. This is the same approach that references \citep{Franco2000,Fattoyev2018} use. We use it in this work because it reduces the computational complexity of the finite-element method significantly. 

We write the perturbed hydro-elastic equilibrium equations for the solid crust in the weak form. The variational weak-form formulation of our problem is given by
\[
    \int_{\Omega_s} \overleftrightarrow{\sigma}
    (\vec{u}):\overleftrightarrow{\varepsilon}(\delta \vec{u}) \ dV \ + \ \int_{\Omega_s} \delta \rho (\vec{u}) \ \vec{\nabla} \Phi \cdot \delta \vec{u} \ dV \\   
\]
\[
    + \ \int_{\Omega_s} \rho \ \vec{\nabla} \delta \Phi(\vec{u}) \cdot \delta \vec{u} \ dV - \int_{\partial \Omega_s} (\overleftrightarrow{\sigma}(\vec{u}) \cdot \hat{n}) \cdot \delta \vec{u} \ dS \\
\]
\begin{equation}
    = \int_{\Omega_s} \vec{f} \cdot \delta \vec{u} \ dV \ ,
    \label{eqn:pert_Euler_crust_weak}
\end{equation}
where : represents an inner product between two tensors. In the variational formulation, the displacement $\vec{u}$ is the trial function that we want to approximate while $\delta \vec{u}$ is the test function. $\overleftrightarrow{\varepsilon}$ is the elastic strain and $\overleftrightarrow{\sigma}$ is the sum of the elastic stress and the pressure. $\Omega_s$ is the integration domain (the solid crust) and $\partial \Omega_s$ is the sum of all the boundaries (the bottom and top of the crust). $\hat{n}$ represents a unit vector that is normal to the boundary surfaces (the bottom and top of the crust). Notice that the fourth term on the left-hand side of Eqn. (\ref{eqn:pert_Euler_crust_weak}) corresponds to the continuity of the radial traction across the bottom and top of the crust. The mesh to which we apply the finite-element method is an octant of a sphere with elements that have a size of approximately 100 m \citep{Morales:2024zka}.


\section{Results}

\begin{figure}[h!tbp]
    \centering
    \includegraphics[width=0.45\textwidth]{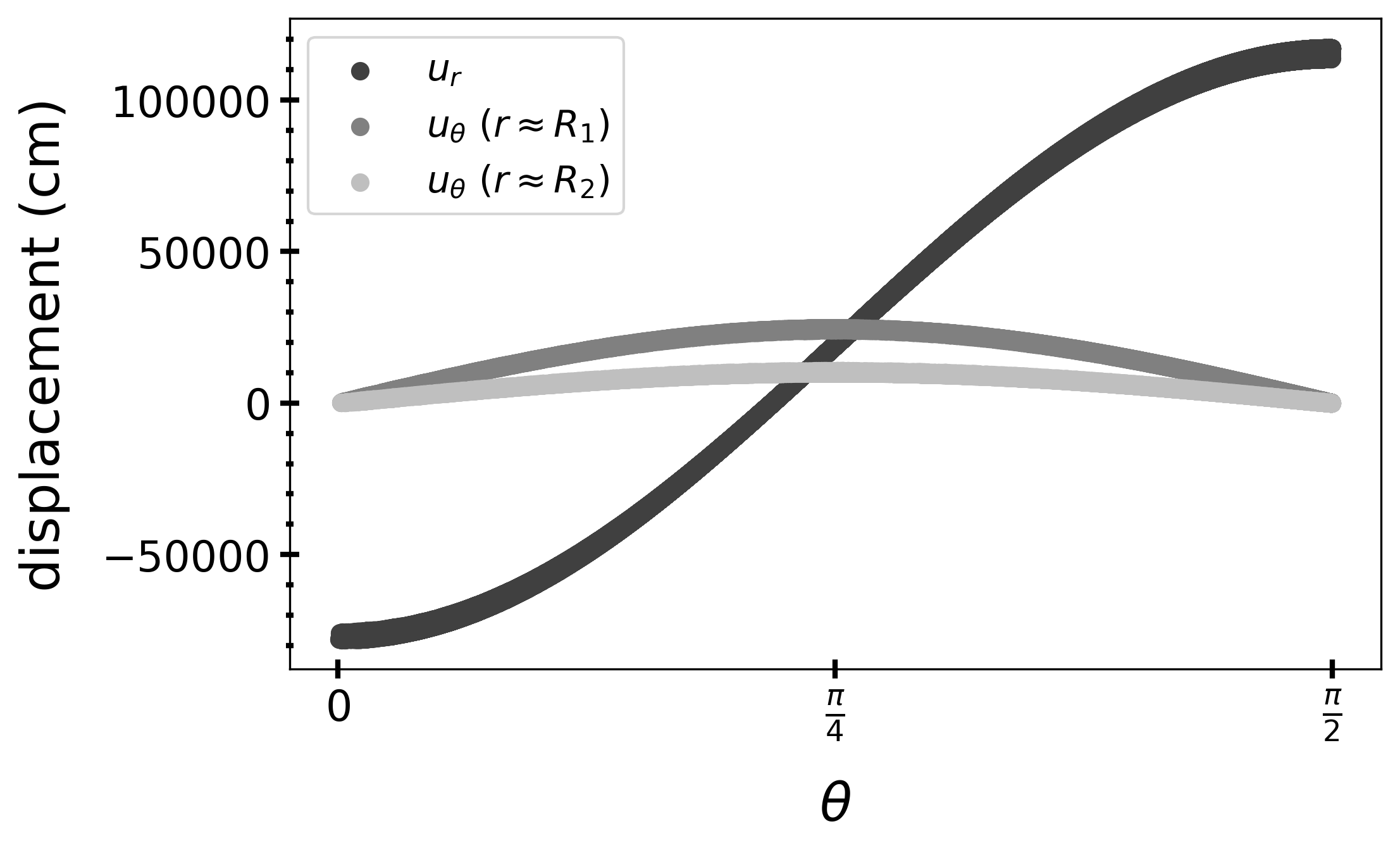}
    \caption{Displacement $u_r$ versus $\theta$  (black). This is almost independent of $r$.  Also shown is  
    $u_\theta$ versus $\theta$ near the bottom of the crust $R_1$ (gray) and near the top of the crust $R_2$ (light gray).}
    \label{fig:u_r_u_th_th} 
\end{figure}

The finite-element method yields the Eulerian displacements. These are shown in Fig.~\ref{fig:u_r_u_th_th} where $u_r$ and $u_\theta$ represent the  $r$ and $\theta$ components of the displacement in spherical coordinates.  This is for a 1.4 $M_\odot$, $R=10$ km star that has been spun up from rest to $\Omega=\Omega_K/2$ where $\Omega_K=(GM/R^3)^{1/2}$ is the (nonrelativistic) Kepler frequency.  Material near the poles $\theta\approx 0$ moves in while material near the equator $\theta\approx\pi/2$ moves out to form the equatorial bulge.

\begin{figure}[h!tbp]
    \centering
\includegraphics[width=0.45\textwidth]{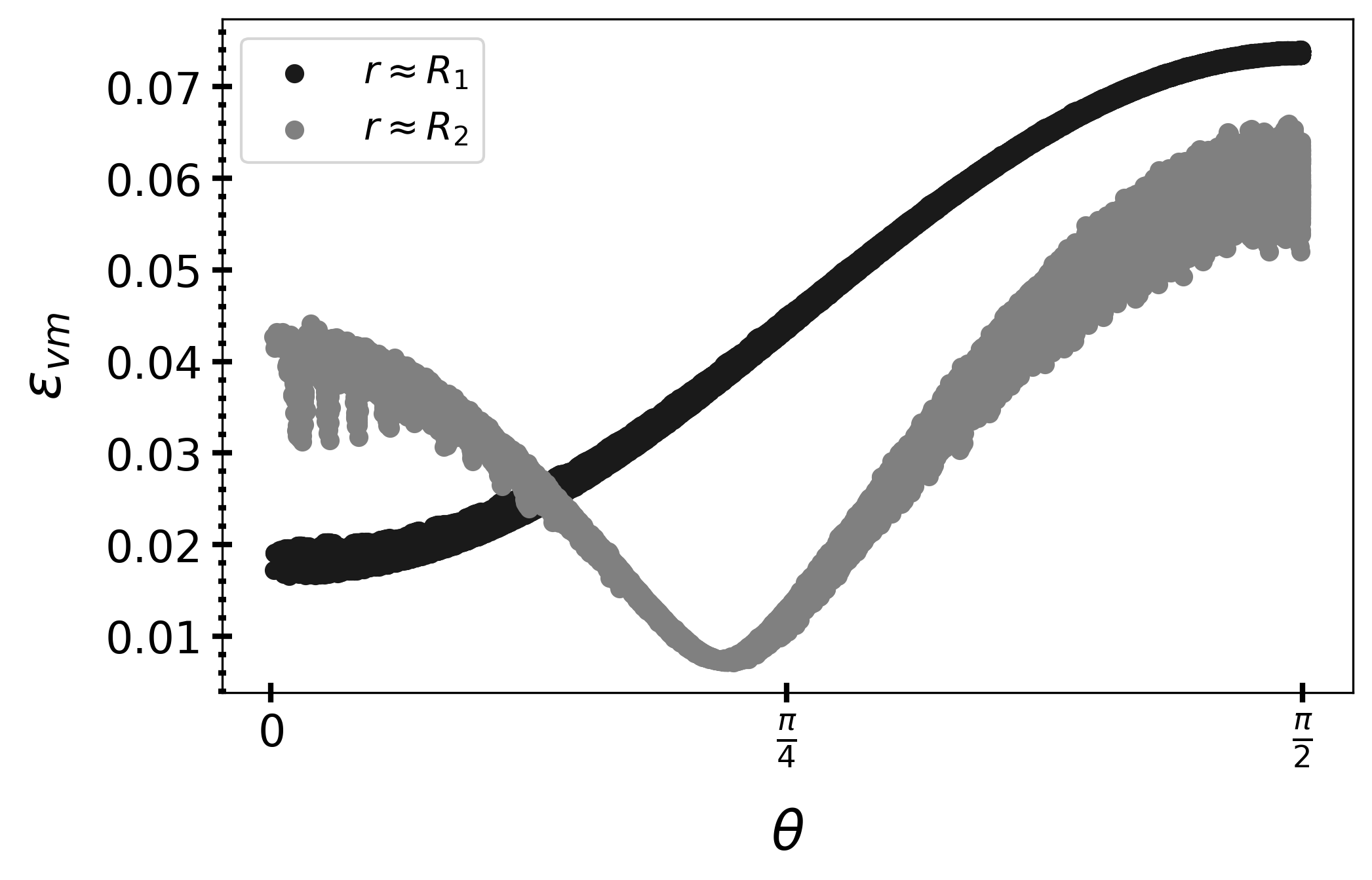}
    \caption{von-Mises strain  versus $\theta$ near the base of the crust ($r=R_1$, black band) and near the top of the crust ($r=R_2$, gray band).}
    \label{fig:strain_th} 
\end{figure}

The strain $\varepsilon_{ij}$ follows from derivatives of the displacement.  In Cartesian coordinates $\varepsilon_{ij}$ is 
\begin{equation}
    \varepsilon_{ij} = \frac{1}{2} \left( \nabla_i u_j + \nabla_j u_i \right)\ .
\end{equation}
This includes deformations due to changes in shape and volume.  We subtract contributions from the change in volume and define the deviatoric strain $\bar{\varepsilon}_{ij}$ that only includes contributions from the change in shape,
\begin{equation}
    \bar{\varepsilon}_{ij} = \varepsilon_{ij} - \frac{1}{3} \delta_{ij} \sum_{k=1}^3\varepsilon_{kk} \ .
    \label{eqn:strain}
\end{equation}
We assume the crust fails when the von-Mises strain exceeds the breaking strain.  The von-Mises strain is defined as
 $   \varepsilon_{vm} = \sqrt{ \frac{1}{2} \overleftrightarrow{\bar{\varepsilon}} : \overleftrightarrow{\bar{\varepsilon}}} $$.
$     
In spherical coordinates, this can be written as
\begin{equation}
    \varepsilon_{vm} = \sqrt{ \frac{1}{2} \left( \bar{\varepsilon}_{rr}^2 + \bar{\varepsilon}_{\theta \theta}^2 + \bar{\varepsilon}_{\phi \phi}^2 + 2 (\bar{\varepsilon}_{r \theta}^2 + \bar{\varepsilon}_{r\phi}^2 + \bar{\varepsilon}_{\theta\phi}^2) \right) }\ .
    \label{eqn:strain_vm_spherical}
\end{equation}
The NS that spins up from rest to $\Omega_K/2$ has the von-Mises strain field shown in Fig. \ref{fig:strain_th}. The von-Mises strain is largest at the base of the crust ($r=R_1$) and at the equator ($\theta=\pi/2$).

\begin{figure}[h!tbp]
    \centering
    \includegraphics[width=0.45\textwidth]{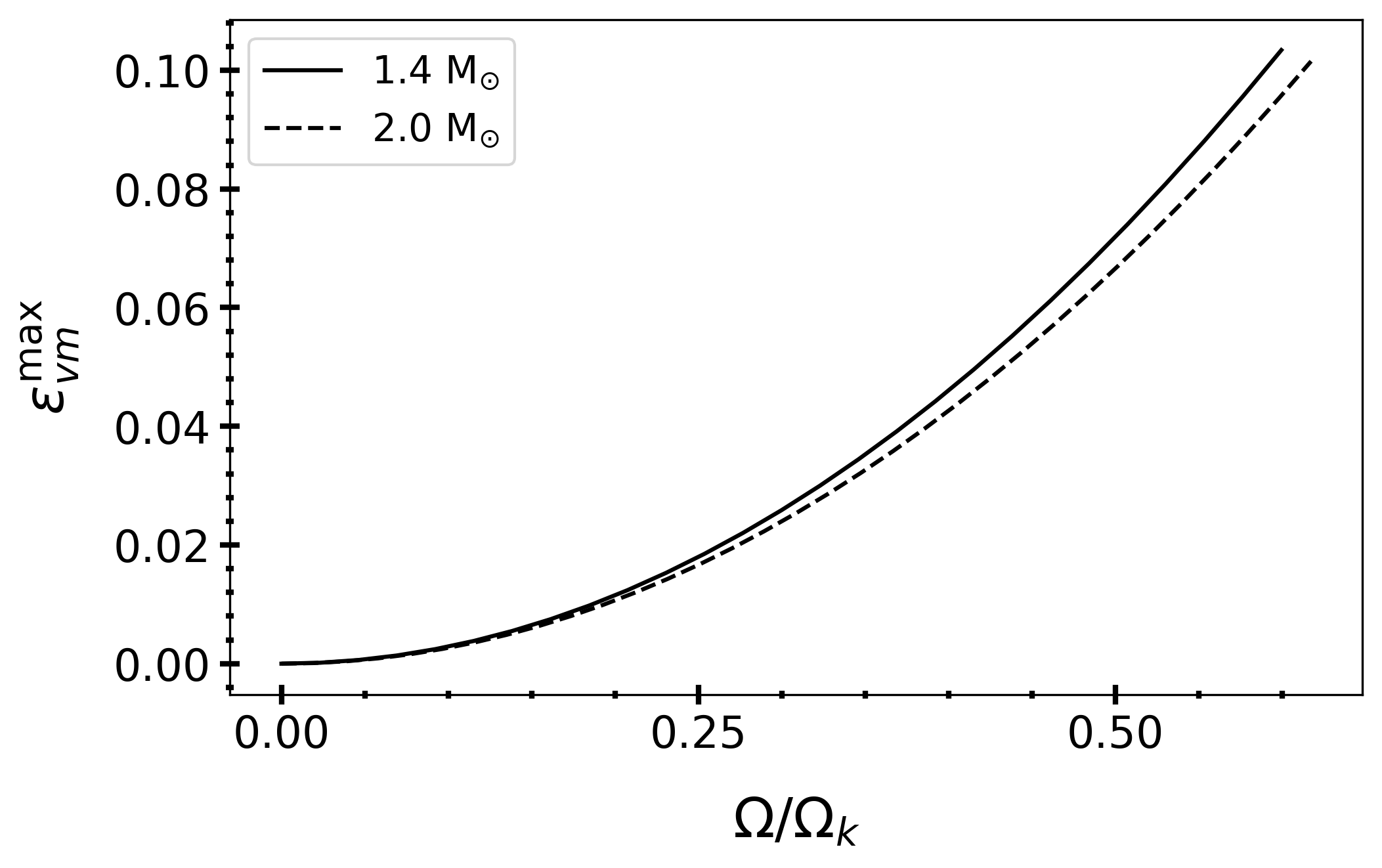}
    \caption{Maximum von-Mises strain versus rotational frequency $\Omega$ in units of the Keplerian (break up) frequency $\Omega_K$.  The solid curve is for a 1.4 M$_\odot$ and the dashed curve for a 2 M$_\odot$ star.}   \label{fig:Omega} 
\end{figure}

We find that the crust breaks first at the equator of the star. This confirms the results that \cite{Fattoyev2018} and \cite{Franco2000} obtained with their simple constant-density NS model. In addition, we find that the crust breaks first at the base ($r=R_1$). This assumes the breaking strain is independent of density. Breaking at the base of the crust is in a high density region where nuclear pasta may be present.  The breaking strain of pasta may be similar to that for the lower density crust \citep{PhysRevLett.121.132701}, however elastic properties of pasta should be studied further.

Figure \ref{fig:Omega} shows that the maximum strain of the crust grows with rotational frequency.  A strain of 0.1 is reached when the star is rotating at 0.58 of the Kepler breakup rate.   This maximum strain is almost independent of the star's mass. Results for a 2 M$_\odot$ star are very similar to those for a 1.4 M$_\odot$ star, assuming the same 10 km radius.  

Our simulations are for Newtonian gravity.  There are many fully relativistic calculations of the shape of rotating {\it fluid} stars, for example \citep{1986ApJ...304..115F,Konstantinou_2022}.  Perhaps the largest relativistic effect is a decrease of about one third from the non-relativistic Kepler frequency $\Omega_K=(GM/R^3)^{1/2}$ to the relativistic Kepler frequency $\Omega^r_K$ \citep{Konstantinou_2022}, see Table~\ref{Table1}. This reduction may also be due in part to an increase in the equatorial radius with increasing spin.

\begin{table}[htbp]
\begin{center}
    
  \begin{tabular}{l | l  l | l l|l l}
    $R$ & $\Omega_K$ & $\Omega^r_K$ & 
    \multicolumn{2}{l|} {${\Delta R_e/R}$}   & \multicolumn{2}{l}{$\varepsilon^{\rm max}_{\rm vm}$} \\
    (km) &(Hz)  &(Hz) & 0.05 & 0.1 & 0.05 & 0.1\\
    \hline
11 & 1890 & 1300 & 690 & 940 & 790 & 1100  \\
12 & 1650 & 1130 &600 & 810 & 690 & 960 \\
13 & 1470 & 1000 & 530 & 720 & 620 & 850 \\
 
    \end{tabular}
    \end{center}\caption{Rotational frequencies for a 1.4M$_\odot$ neutron star of radius $R$.  The non-relativistic (relativistic) Kepler frequency is $\Omega_K$ ($\Omega^r_K)$ . The columns $\Delta R_e/R=0.05(0.1)$ give the frequency (in Hz) where the equatorial radius of a fluid star increases by 5\%(10\%) \citep{Konstantinou_2022}.  Finally, the last two columns give the frequency (in Hz) where the crust fails in our simulations for a breaking strain of 0.05 or 0.1.   } \label{Table1} 
\end{table}

We use a simple polytropic equation of state. \cite{Konstantinou_2022} find nearly universal relations where the equation of state dependence is primarily through the star's radius $R$ (via the $R^{-3/2}$ dependence of $\Omega_K$).  Given NS radii determinations from X-ray and gravitational wave observations and nuclear experimental and theoretical information \citep{galaxies10050099,multiNSradii,HICNSradii}, we estimate equation of state uncertainties by considering $R$ from 11 to 13 km in Table~\ref{Table1}.   Perhaps the largest uncertainty is the exact value of the breaking strain.  Molecular dynamics simulations predict that the breaking strain is {\it of order} 0.1.  However the exact value could be somewhat different.  For example a breaking strain of 0.05 \citep{10.1093/mnras/sty2259} is reached at 0.42 of the breakup speed.  

Finally as a ``sanity check'' and to test for large relativistic effects we estimate crustal strain from relativistic calculations of the change in radius with spin of {\it fluid} stars.  Qualitatively the crustal strain $\varepsilon_{rr}$ is of order $\Delta R_e/R$ where $\Delta R_e$ is the increase in equatorial radius with spin of a fluid star.  Table~\ref{Table1} lists the spin frequencies where $R_e$ increases by 5\% or 10\% \citep{Konstantinou_2022}.  These values are similar to but slightly smaller than our non-relativistic simulation results.  This suggests that there is not a large error in our simulations and relativistic effects are not large.  

Our simulations were performed for $R=10$ km.  We scale our results to other radii by assuming $\Omega$ scales with $\Omega_K(R)$ because \cite{Konstantinou_2022} find $\Delta R_e/R$ is approximately independent of $R$ for fixed $\Omega/\Omega_K^r$.  This yields the $\Omega$ values in Table~\ref{Table1}.  For example, we find the crust fails at $\Omega/\Omega_K=0.42$ assuming the breaking strain is 0.05.  The nonrelativistic Kepler frequency of a $R=12$ km star is 1650 Hz.  Therefore, we predict that the crust of a $R=12$ km 1.4M$_\odot$ star will fail at $\Omega=0.42\times 1690=690$ Hz. 
This frequency is comparable to the 716 Hz rotation frequency of the fastest known NS \citep{2006Sci...311.1901H}.  However, Table~\ref{Table1} indicates significant uncertainties.

\section{Discussion}

When the crust breaks, the release of elastic stresses will allow parts of the crust to move.  This could produce a nonzero ellipticity $\epsilon$.  This is a measure of the asymmetry of the star,
\begin{equation}
  \epsilon=\frac{I_1-I_2}{I_3}\,.  
\end{equation}
Here $I_1,I_2$, and $I_3$ are the principle moments of inertia and the star is rotating about the 3 axis.  The ellipticity is the important parameter to determine the characteristic strain amplitude $h_0$ of GW radiation from a star with rotational frequency $\Omega$ at a distance $d$ \citep{Riles2023},
\begin{equation}
    h_0 = \frac{16 \pi^2 G}{c^4} \frac{I_3}{d} \Omega^2 \epsilon \ .
    \label{eqn:intrinsic_strain}
\end{equation}
Here $G$ is Newton's constant and $c$ is the speed of light.

We have not attempted to model what happens to the crust as it fails and what the final crust configuration might be.  This may require many assumptions about crust breaking microphysics.  Therefore, we do not directly calculate $\epsilon$.   However, we now argue $\epsilon$ is likely to be significant $\ge 10^{-8}$ because of the following considerations.
\begin{enumerate}
    \item 
Much of the crust is strained to near (or at) the breaking strain so that deformations of the crust are large.  
\item 
The maximum ellipticity the crust can support $\epsilon_{max}\approx 10^{-5}$\citep{10.1093/mnras/stac3058, PhysRevLett.102.191102} is much larger that the $\epsilon\approx 10^{-8}$ that may be needed for torque balance.  Before crust failure the deformations are axisymmetric and $\epsilon=0$.  If crust failure releases strain on one side of the star, the remaining deformations may have a significant quadrupole pattern so that $\epsilon$ is nonzero and could be a noticeable fraction of $\epsilon_{max}$.  

\item
The first failures occur near the base of the crust where the density is large.  This region involves a significant fraction of the mass of the crust and can source a large ellipticity.
\end{enumerate}

If $\epsilon\ge 10^{-8}$ after crust breaking, the system may rapidly reach torque equilibrium between GW radiation and accretion \citep{10.1093/mnras/stad2036} because the GW torque $\propto \Omega^5$ is a stiff function of the rotation rate.  If $\epsilon$ is initially somewhat too large the star will spin down slightly from GW radiation until it reaches torque equilibrium.  Alternatively if $\epsilon$ is somewhat too small the system may spin up slightly from accretion until it reaches equilibrium.  In either case the system will reach torque equilibrium at a rotation rate near where the crust failed.  Therefore, {\it crust breaking may determine the maximum rotation rate of neutron stars.}

An accreting star can have a thick ocean and a magnetic field threading the crust that can respond to crust failure.  Surface effects at low densities will likely not contribute significantly to $\epsilon$.  However, they may produce electromagnetic signals associated with crust breaking, see for example \citep{Tsang_2013}.

We expect $\epsilon$ to evolve with time.  Further crust failures, depending on where they are located, can either increase or decrease $\epsilon$.  Elastic stress and $\epsilon$ can decrease with time due to  viscoelastic creep \citep{10.1111/j.1745-3933.2010.00903.x}.   Further accretion can also decrease $\epsilon$ as crust material is buried to densities beyond the crust-core transition and melted. These effects could reduce $\epsilon$ to $\approx 10^{-9}$ after accretion stops.  Woan et al argue that the observed spin down rates of millisecond pulsar populations suggest such a residual ellipticity \citep{Woan_2018}. 

A population of rapidly rotating NS with significant ellipticity from crust breaking could be very promising sources for continuous GW searches \citep{Pagliaro2023,Reed_2021}.  In general, rapidly rotating stars will produce stronger GW because of the factor of $\Omega^2$ in Eq.~\ref{eqn:intrinsic_strain}.  Furthermore, search sensitivity will improve with improved techniques and with better next generation detectors. 

In conclusion,  neutron stars (NS) are observed to spin no faster than about half the breakup rate.  This may be related to the strength of their crusts.   We performed finite-element simulations of rotating neutron stars and find the crust fails at rotation rates about half the breakup rate.   This crust failure could produce an asymmetric NS with a significant ellipticity (fractional difference in moments of inertia).  Rapidly rotating stars with this ellipticity radiate gravitational waves that could limit further spin up and may be promising sources for LIGO / VIRGO and next generation detectors.

\

Acknowledgements: We thank Cole Miller and Gianluca Pagliaro for helpful comments. We thank the developers of \texttt{FEniCSx} for their help during the development and implementation of our finite-element simulations. This work is partially supported by the US Department of Energy grant DE-FG02-87ER40365 and National Science Foundation grant PHY-2116686. 




\begin{thebibliography}{}
\expandafter\ifx\csname natexlab\endcsname\relax\def\natexlab#1{#1}\fi
\providecommand{\url}[1]{\href{#1}{#1}}
\providecommand{\dodoi}[1]{doi:~\href{http://doi.org/#1}{\nolinkurl{#1}}}
\providecommand{\doeprint}[1]{\href{http://ascl.net/#1}{\nolinkurl{http://ascl.net/#1}}}
\providecommand{\doarXiv}[1]{\href{https://arxiv.org/abs/#1}{\nolinkurl{https://arxiv.org/abs/#1}}}

\bibitem[{Abbott {et~al.}(2017{\natexlab{a}})}]{PhysRevD.95.122003}
Abbott, B.~P., {et~al.} 2017{\natexlab{a}}, Phys. Rev. D, 95, 122003,
  \dodoi{10.1103/PhysRevD.95.122003}

\bibitem[{Abbott {et~al.}(2017{\natexlab{b}})Abbott, Abbott, Abbott, Abraham,
  Acernese, Ackley, Adams, Adhikari, Adya, \& Affeldt}]{Abbott2017_4}
Abbott, B.~P., Abbott, R., Abbott, T.~D., {et~al.} 2017{\natexlab{b}}, ApJ,
  847, 47, \dodoi{10.3847/1538-4357/aa86f0}

\bibitem[{Abbott {et~al.}(2019)Abbott, Abbott, Abbott, Abraham, Acernese,
  Ackley, Adams, Adhikari, Adya, \& Affeldt}]{Abbott2019_3}
---. 2019, Phys. Rev. D, 100, 122002, \dodoi{10.1103/PhysRevD.100.122002}

\bibitem[{Abbott {et~al.}(2022{\natexlab{a}})Abbott, Abbott, Acernese, Ackley,
  Adams, Adhikari, Adhikari, \& Adya}]{Abbott2022_3}
Abbott, R., Abbott, T.~D., Acernese, F., {et~al.} 2022{\natexlab{a}}, Phys.
  Rev. D, 105, 022002, \dodoi{10.1103/PhysRevD.105.022002}

\bibitem[{Abbott {et~al.}(2022{\natexlab{b}})}]{Abbott_2022new}
Abbott, R., {et~al.} 2022{\natexlab{b}}, The Astrophysical Journal Letters,
  941, L30, \dodoi{10.3847/2041-8213/aca1b0}

\bibitem[{Andersson {et~al.}(2005)Andersson, Glampedakis, Haskell, \&
  Watts}]{10.1111/j.1365-2966.2005.09167.x}
Andersson, N., Glampedakis, K., Haskell, B., \& Watts, A.~L. 2005, Monthly
  Notices of the Royal Astronomical Society, 361, 1153,
  \dodoi{10.1111/j.1365-2966.2005.09167.x}

\bibitem[{Baiko \& Chugunov(2018)}]{10.1093/mnras/sty2259}
Baiko, D.~A., \& Chugunov, A.~I. 2018, Monthly Notices of the Royal
  Astronomical Society, 480, 5511, \dodoi{10.1093/mnras/sty2259}

\bibitem[{Baym \& Pines(1971)}]{BAYM1971816}
Baym, G., \& Pines, D. 1971, Annals of Physics, 66, 816,
  \dodoi{https://doi.org/10.1016/0003-4916(71)90084-4}

\bibitem[{Bildsten(1998)}]{Bildsten1998}
Bildsten, L. 1998, The Astrophysical Journal, 501, L89, \dodoi{10.1086/311440}

\bibitem[{Caplan {et~al.}(2018)Caplan, Schneider, \&
  Horowitz}]{PhysRevLett.121.132701}
Caplan, M.~E., Schneider, A.~S., \& Horowitz, C.~J. 2018, Phys. Rev. Lett.,
  121, 132701, \dodoi{10.1103/PhysRevLett.121.132701}

\bibitem[{Chugunov \& Horowitz(2010)}]{10.1111/j.1745-3933.2010.00903.x}
Chugunov, A.~I., \& Horowitz, C.~J. 2010, Monthly Notices of the Royal
  Astronomical Society: Letters, 407, L54,
  \dodoi{10.1111/j.1745-3933.2010.00903.x}

\bibitem[{Fattoyev {et~al.}(2018)Fattoyev, Horowitz, \& Lu}]{Fattoyev2018}
Fattoyev, F.~J., Horowitz, C.~J., \& Lu, H. 2018, arXiv:1804.04952

\bibitem[{Franco {et~al.}(2000)Franco, Link, \& Epstein}]{Franco2000}
Franco, L.~M., Link, B., \& Epstein, R.~I. 2000, The Astrophysical Journal,
  543, 987, \dodoi{10.1086/317121}

\bibitem[{{Friedman} {et~al.}(1986){Friedman}, {Ipser}, \&
  {Parker}}]{1986ApJ...304..115F}
{Friedman}, J.~L., {Ipser}, J.~R., \& {Parker}, L. 1986, \apj, 304, 115,
  \dodoi{10.1086/164149}

\bibitem[{Haskell {et~al.}(2006)Haskell, Jones, \& Andersson}]{Haskell2006}
Haskell, B., Jones, D.~I., \& Andersson, N. 2006, Monthly Notices of the Royal
  Astronomical Society, 373, 1423, \dodoi{10.1111/j.1365-2966.2006.10998.x}

\bibitem[{Haskell \& Patruno(2011)}]{Haskell_2011}
Haskell, B., \& Patruno, A. 2011, The Astrophysical Journal Letters, 738, L14,
  \dodoi{10.1088/2041-8205/738/1/L14}

\bibitem[{{Hessels} {et~al.}(2006){Hessels}, {Ransom}, {Stairs}, {Freire},
  {Kaspi}, \& {Camilo}}]{2006Sci...311.1901H}
{Hessels}, J. W.~T., {Ransom}, S.~M., {Stairs}, I.~H., {et~al.} 2006, Science,
  311, 1901, \dodoi{10.1126/science.1123430}

\bibitem[{Horowitz \& Kadau(2009)}]{PhysRevLett.102.191102}
Horowitz, C.~J., \& Kadau, K. 2009, Phys. Rev. Lett., 102, 191102,
  \dodoi{10.1103/PhysRevLett.102.191102}

\bibitem[{Huth {et~al.}(2022)Huth, Pang, Tews, Dietrich, Le~F{\`e}vre, Schwenk,
  Trautmann, Agarwal, Bulla, Coughlin, \& Van Den~Broeck}]{HICNSradii}
Huth, S., Pang, P. T.~H., Tews, I., {et~al.} 2022, Nature, 606, 276,
  \dodoi{10.1038/s41586-022-04750-w}

\bibitem[{Jones \& Hutchins(2024)}]{jones2024neutronstarmountainssupported}
Jones, D.~I., \& Hutchins, T.~J. 2024, Neutron star mountains supported by
  crustal lattice pressure.
\newblock \doarXiv{2407.00162}

\bibitem[{Konstantinou \& Morsink(2022)}]{Konstantinou_2022}
Konstantinou, A., \& Morsink, S.~M. 2022, The Astrophysical Journal, 934, 139,
  \dodoi{10.3847/1538-4357/ac7b86}

\bibitem[{Lim \& Holt(2022)}]{galaxies10050099}
Lim, Y., \& Holt, J.~W. 2022, Galaxies, 10, \dodoi{10.3390/galaxies10050099}

\bibitem[{{Manchester} {et~al.}(2005){Manchester}, {Hobbs}, {Teoh}, \&
  {Hobbs}}]{2005AJ....129.1993M}
{Manchester}, R.~N., {Hobbs}, G.~B., {Teoh}, A., \& {Hobbs}, M. 2005, The
  Astrophysical Journal, 129, 1993, \dodoi{10.1086/428488}

\bibitem[{Morales \& Horowitz(2022)}]{10.1093/mnras/stac3058}
Morales, J.~A., \& Horowitz, C.~J. 2022, Monthly Notices of the Royal
  Astronomical Society, 517, 5610, \dodoi{10.1093/mnras/stac3058}

\bibitem[{Morales \& Horowitz(2024)}]{Morales:2024zka}
---. 2024, 2409.14482

\bibitem[{Pagliaro {et~al.}(2023)Pagliaro, Papa, Ming, Lian, Tsuna, Maraston,
  \& Thomas}]{Pagliaro2023}
Pagliaro, G., Papa, M.~A., Ming, J., {et~al.} 2023, The Astrophysical Journal,
  952, 123, \dodoi{10.3847/1538-4357/acd76f}

\bibitem[{Pang {et~al.}(2023)Pang, Dietrich, Coughlin, Bulla, Tews, Almualla,
  Barna, Kiendrebeogo, Kunert, Mansingh, Reed, Sravan, Toivonen, Antier,
  VandenBerg, Heinzel, Nedora, Salehi, Sharma, Somasundaram, \& Van
  Den~Broeck}]{multiNSradii}
Pang, P. T.~H., Dietrich, T., Coughlin, M.~W., {et~al.} 2023, Nature
  Communications, 14, 8352, \dodoi{10.1038/s41467-023-43932-6}

\bibitem[{Parfrey {et~al.}(2016)Parfrey, Spitkovsky, \&
  Beloborodov}]{Parfrey_2016}
Parfrey, K., Spitkovsky, A., \& Beloborodov, A.~M. 2016, The Astrophysical
  Journal, 822, 33, \dodoi{10.3847/0004-637X/822/1/33}

\bibitem[{{Radhakrishnan} \& {Srinivasan}(1982)}]{1982CSci...51.1096R}
{Radhakrishnan}, V., \& {Srinivasan}, G. 1982, Current Science, 51, 1096

\bibitem[{Reed {et~al.}(2021)Reed, Deibel, \& Horowitz}]{Reed_2021}
Reed, B.~T., Deibel, A., \& Horowitz, C.~J. 2021, The Astrophysical Journal,
  921, 89, \dodoi{10.3847/1538-4357/ac1c04}

\bibitem[{Reitze {et~al.}(2019)Reitze, Adhikari, Ballmer, Barish, Barsotti,
  Billingsley, Brown, Chen, Coyne, Eisenstein, Evans, Fritschel, Hall,
  Lazzarini, Lovelace, Read, Sathyaprakash, Shoemaker, Smith, Torrie, Vitale,
  Weiss, Wipf, \& Zucker}]{reitze2019cosmicexploreruscontribution}
Reitze, D., Adhikari, R.~X., Ballmer, S., {et~al.} 2019, Cosmic Explorer: The
  U.S. Contribution to Gravitational-Wave Astronomy beyond LIGO.
\newblock \doarXiv{1907.04833}

\bibitem[{Riles(2023)}]{Riles2023}
Riles, K. 2023, Living Reviews in Relativity, 26, 3,
  \dodoi{10.1007/s41114-023-00044-3}

\bibitem[{{Ruderman}(1976)}]{1976ApJ...203..213R}
{Ruderman}, M. 1976, The Astrophysical Journal, 203, 213,
  \dodoi{10.1086/154069}

\bibitem[{{Ruderman}(1991)}]{1991ApJ...382..587R}
---. 1991, The Astrophysical Journal, 382, 587, \dodoi{10.1086/170745}

\bibitem[{Tsang(2013)}]{Tsang_2013}
Tsang, D. 2013, The Astrophysical Journal, 777, 103,
  \dodoi{10.1088/0004-637X/777/2/103}

\bibitem[{Ushomirsky {et~al.}(2000)Ushomirsky, Cutler, \&
  Bildsten}]{10.1046/j.1365-8711.2000.03938.x}
Ushomirsky, G., Cutler, C., \& Bildsten, L. 2000, Monthly Notices of the Royal
  Astronomical Society, 319, 902, \dodoi{10.1046/j.1365-8711.2000.03938.x}

\bibitem[{Woan {et~al.}(2018)Woan, Pitkin, Haskell, Jones, \&
  Lasky}]{Woan_2018}
Woan, G., Pitkin, M.~D., Haskell, B., Jones, D.~I., \& Lasky, P.~D. 2018, The
  Astrophysical Journal Letters, 863, L40, \dodoi{10.3847/2041-8213/aad86a}

\bibitem[{Zhang {et~al.}(2021)Zhang, Papa, Krishnan, \& Watts}]{Zhang2021}
Zhang, Y., Papa, M.~A., Krishnan, B., \& Watts, A.~L. 2021, The Astrophysical
  Journal Letters, 906, L14, \dodoi{10.3847/2041-8213/abd256}

\bibitem[{Çıkıntoğlu \& Ekşi(2023)}]{10.1093/mnras/stad2036}
Çıkıntoğlu, S., \& Ekşi, K.~Y. 2023, Monthly Notices of the Royal
  Astronomical Society, 524, 4899, \dodoi{10.1093/mnras/stad2036}

\end{thebibliography}

\end{document}